# Second-order phase transition of silicon from a band insulator to metal induced by strong magnetic fields

Katsuhiko Higuchi[1], Dipendra Bahadur Hamal[2] and Masahiko Higuchi[3]

[1]Graduate School of Advance Sciences of Matter, Hiroshima University, Higashi-Hiroshima 739-8527, Japan
[2]Department of Natural Sciences (Physics), Kathmandu University, Dhulikhel, Kavre 6250, Nepal
[3]Departments of Physics, Faculty of Science, Shinshu University, Matsumoto 390-8621, Japan

**Abstract**

We present the second-order phase transition from a band insulator to metal that is induced by a strong magnetic field. The magnetic-field dependences of the magnetization and energy band gap of a crystalline silicon immersed in a magnetic field are investigated by means of the nonperturbative magnetic-field-containing relativistic tight-binding approximation method [Phys. Rev. B **97**, 195135 (2018)]. It is shown that the energy band gap disappears at the critical magnetic field of $2.22 \times 10^4$ (T). At the critical magnetic field, the magnetic-field dependence of the magnetization exhibits a kink behavior, which means that this phenomenon is the second-order phase transition from a band insulator to metal. It is found that in strong magnetic fields above the critical magnetic field, namely in the metallic phase, the oscillation of the magnetization appears. It is shown that this magnetic oscillation comes from the energy bands in the magnetic Brillouin zone that change from the occupied states to unoccupied states or vice vasa.

# I. Introduction

Materials exhibit various properties in a magnetic field [1-3]. For example, Pauli paramagnetism and Landau diamagnetism that are described on the basis of the free electron model [1-3], Curie paramagnetism and Langevin diamagnetism that are described on the base of the localized electron model [1-3], de Haas-van Alphen (dHvA) effect [4,5] and Shubnikov-de Haas effect [6] are observed in materials immersed in a magnetic field. The quantum Hall effect is observed in two-dimensional electron systems at low temperatures and under a strong magnetic field [7-9]. More recently, an anomalous value of the electron g-factor, which can be explained as the Rashba effect caused by the antisymmetric surface potential, is measured in Graphene by the electron spin resonance experiments [10-12]. Thus, various physical properties are observed in materials immersed in a magnetic field, which are different from those of materials in the zero magnetic field.

There are several kinds of methods to describe electronic states of materials immersed in a magnetic field. For example, the effective mass approximation method [1-3], the semiclassical approach [1-3] and the Hofstadter method [13] are widely used to describe electronic states of materials immersed in a magnetic field. Recently, we developed the magnetic-field-containing relativistic tight-binding approximation (MFRTB) method [14]. This method enables us to calculate the electronic structure of materials immersed in a magnetic field by taking effects of magnetic field, periodic potential and relativity into consideration. This method is applied to a crystalline silicon immersed in the magnetic field parallel to z-axis. It was shown that nearly flat bands are obtained in the $k_x - k_y$ plane of the magnetic first Brillouin zone (BZ), which means that the motion of electrons in the plane perpendicular to the magnetic field is essentially changed corresponding to the quantization of the orbital motion of electrons in a magnetic field [14]. Furthermore, this method is applied to a simple cubic lattice model immersed in a magnetic field. It is shown that one can not only revisit the dHvA oscillation and magnetic breakdown phenomena, but also predict additional oscillation peaks of the magnetization [15,16] that cannot be explained by the conventional Lifshitz-Kosevich formula [17].

In the MFRTB method [14] as well as in the Hofstadter method [14], the magnetic hopping integral is approximately given by the hopping integral in the absence of a magnetic field multiplied by the so-called Peierls phase factor. It is revealed in the preceding work that this approximation for the magnetic hopping integral is obtained within the lowest-order perturbation theory [18]. In order to increase the accuracy of the estimation of magnetic hopping integrals and make the method applicable to materials immersed in a strong magnetic field, the nonperturbative MFRTB method has been developed [18,19]. In nonperturbative MFRTB method, the effect of the magnetic field is incorporated by using a variational method. As a result, we have successfully obtained the approximate form of magnetic hopping integrals that goes beyond the approximation of using the Peierls phase factor and is applicable to materials immersed in a strong magnetic field.

In this paper, it is shown by using the nonperturbative MFRTB method that the second-order phase transition from a band insulator to metal is induced by a strong magnetic field. Specifically, we apply the nonperturbative MFRTB method to a crystalline silicon immersed in a strong magnetic field in order to investigate the magnetic-field dependences of the magnetization and energy band gap of a crystalline silicon. As shown later, the magnetic-field dependence of the magnetization exhibits a kink behavior at a critical magnetic field, at which the energy band gap disappears. It is also shown that the magnetization oscillation appears in strong magnetic fields above the critical magnetic field.

## II. Method

The nonperturbative MFRTB method [18] is applied to the crystalline silicon immersed in a magnetic field. In the calculation, we consider magnetic hopping integrals between the outer shells of nearest-neighbor silicon atom. That is the s- and p-orbitals of the outer shell of the silicon atom (8 orbitals in total) are taken into consideration. We adopt a set of the tight-binding parameters of the crystalline silicon that are given in the previous paper [14]. The magnetic field is supposed to be parallel to z-axis. The magnitude of the magnetic field is given by $B = \left(16\pi\hbar/ea^2\right)(p/q)$, where $a$ denotes the lattice constant of silicon, and where $p$ and $q$ are relatively prime integers [14]. In this case, there are $2q$ silicon atoms in the magnetic primitive unit cell. The $16q$ energy bands are obtained in the magnetic BZ. Since the $\boldsymbol{k}$-dependence of $E(\boldsymbol{k})$ becomes nearly flat in the plane ($k_x - k_y$ plane) perpendicular to the magnetic field [14], we calculate the total energy under the assumption that $E(\boldsymbol{k})$ in the $k_x - k_y$ plane can be approximated by the value of $E(\boldsymbol{k})$ at $(0,0,k_z)$ similarly to the previous works [19,20]. The magnetization is calculated by the derivative of the total energy with respect to the magnetic field.

## III. Results and discussion

Figure 1 shows the magnetic-field dependences of the magnetization and energy band gap. The magnetic-field dependence of the magnetization exhibits a kink behavior at the magnetic field of $B_c = 2.22 \times 10^4$ (T). Namely, the magnetization increases monotonically in magnetic fields below $B_c$. The overall slope of the magnetization significantly changes at $B_c$. The magnetization that corresponds to first derivative of the total energy with respect to $B$ is continuous across $B_c$, but the second derivative exhibits discontinuity. This suggests that this phenomenon is regarded as the second-order phase transition induced by the magnetic field.

As shown in Fig. 1, the energy band gap monotonically increases with $B$ up to about $5 \times 10^3$ (T). In magnetic fields above $5 \times 10^3$ (T), the energy band gap shows a decreasing trend with increasing $B$, and eventually reaches zero at $B_c = 2.22 \times 10^4$ (T). This means that the crystalline silicon in strong magnetic fields above $B_c$ changes to metal. Therefore, we can say that the crystalline silicon exhibits the second-order phase transition from a band insulator to metal induced by the magnetic field.

Let us consider this second-order phase transition in more detail by using the energy band structure obtained in the magnetic BZ. Figure 2 shows $E(\mathbf{k})$ for $B = 743.1$ (T) that is much smaller than $B_c$. The horizontal axes denotes the normalized wavevector $\bar{\mathbf{k}}$ that is defined as $\mathbf{k} = \frac{2\pi}{a} \bar{\mathbf{k}}$. In Fig. 2, $\bar{\mathbf{k}}$ changes from $\Gamma$ point ($\bar{\mathbf{k}} = (0,0,0)$) and returns to $\Gamma$ point via $\bar{\mathbf{k}} = (0,0,0.5)$ and $R_x$ point ($\bar{\mathbf{k}} = (1,0,0)$). Similar to the MFRTB calculations [14], the nonperturbed MFRTB method also yields nearly flat energy bands in the $k_x - k_y$ plane as shown in Fig. 2. The highest occupied state and lowest unoccupied state are obtained at $\Gamma$ point and $\bar{\mathbf{k}} = (0,0,0.5)$, respectively. The energy band gap at $B = 743.1$ (T) is about 1.504 (eV) that is larger than that for the case of zero magnetic field.

Figures 3(a), 3(b) and 3(c) show the energy bands at (a) $B = 0.9999 \times 10^4$ (T), (b) $B = 1.5554 \times 10^4$ (T) and (c) $B = 2.2156 \times 10^4$ (T), respectively, that are lower than $B_c$. Figure 4(d) shows energy bands at $B = 2.2220 \times 10^4$ (T) that is higher than $B_c$. The horizonal axes denotes $\bar{\mathbf{k}}$ that changes from $\Gamma$ point to $R_z$ point ($\bar{\mathbf{k}} = (0,0,1)$). As shown in Figs. 3(a) - 3(c), the normalized wavevectors that gives the highest occupied and lowest unoccupied states shift away from those for $B = 743.1$ (T) (Fig. 2) with increasing $B$. In the magnetic field just before the energy band gap disappears (Fig. 3(c)), the highest occupied states are obtained at $\bar{\mathbf{k}} = (0,0,0.447)$ and $(0,0,0.553)$, while the lowest unoccupied states are obtained at $\bar{\mathbf{k}} = (0,0,0.478)$ and $\bar{\mathbf{k}} = (0,0,0.522)$. The energy band gap is about $7.89 \times 10^{-3}$ (eV) at $B = 2.2156 \times 10^4$ (T). The energy band gap disappears at $B = 2.2220 \times 10^4$ (T) as shown in Fig. 3(d).

In the metallic phase ($B > B_c$), the magnetization increases in oscillation with increasing $B$. Since the crystalline silicon is in the metallic phase at strong magnetic fields above $B_c$, the appearance of magnetic oscillation seems to be reasonable. However, it should be notice that there does not exist the Fermi surface for silicon at the zero magnetic field, so that this magnetic oscillation is different from the dHvA oscillation observed in usual metals. The present magnetic oscillations are caused by energy states changing from occupied to unoccupied states with increasing $B$. Indeed, we confirm this description of the magnetic oscillation from the energy bands obtained in the magnetic BZ. Let us focus on the oscillation peak observed around

$B = 2.5 \times 10^4$ (T) in Fig. 1. Figures 4(a), 4(b), 4(c) and 4(d) show the energy bands at $B = 2.4725 \times 10^4$ (T), $B = 2.4826 \times 10^4$ (T), $B = 2.5015 \times 10^4$ (T) and $B = 2.5169 \times 10^4$ (T), respectively. As shown in Fig. 4(a), a set of unoccupied energy bands exist near $\bar{k} = (0,0,0.5)$ in the energy ranging from -6.57 (eV) to -6.55 (eV). Due to these energy states, the density of states becomes much large. This is because in the plane of $\bar{k}_z = 0.5$ there exist a lot of states that have almost the same energy as that with $\bar{k} = (0,0,0.5)$. As shown in Figs. 4(b), these energy bands move down with increasing $B$. These energy bands move down with increasing $B$ and eventually become occupied states in Figs. 4(c) and 4(d). Corresponding to this change, the magnetization oscillates around $B = 2.5 \times 10^4$ (T) as shown in Fig. 1.

While the magnetization oscillates with increasing $B$ in magnetic fields above $B_c$, the magnetization increases without oscillations in magnetic fields below $B_c$ as shown in Fig. 1. We shall give a comment on this point. The number of $k$-points in the magnetic BZ is equal to that of the magnetic primitive unit cell contained in the system [14]. The number of the magnetic primitive unit cell is given as $N/q$, where $N$ denotes the number of the primitive unit cells in the zero magnetic field case [14,21]. Therefore, the number of $k$-points in the magnetic BZ is given by $N/q$, which means that each energy band contains $N/q$ electronic states. Since there are $8N$ valence electrons in the system, these electrons will occupy the states of the lower $8q$ energy bands. Even if the magnetic field changes, the lower $8q$ energy bands are occupied until the energy gap disappears. It is shown in the previous work [15] that oscillations in the magnetization such as the dHvA oscillation and additional oscillation peaks [15] occur when energy bands change from occupied states to unoccupied ones or vice vasa. Since the lower $8q$ energy bands are always occupied when an energy gap exists, energy bands never change from occupied states to unoccupied states or vice vasa. Therefore, oscillations in the magnetization would not occur in the band insulator phase.

**IV. Concluding remarks**

In summary, it is found that the second-order phase transition is induced in the crystalline silicon by a strong magnetic field. The critical magnetic field is about $2.22 \times 10^4$ (T). The energy band gap disappears at the critical magnetic field, which means that this transition is recognized as a band-insulator to metal transition induced by the magnetic field. It is also found that a magnetic oscillation in the magnetization appears after the transition. This is because the energy bands in the magnetic BZ change from the occupied to unoccupied states or vice vasa.

We shall give a brief comment on the magnitude of the critical magnetic field. In this work, we have treated the second-order phase transition of the crystalline silicon that has an energy band gap of about 1.1 (eV) at zero magnetic field. For materials with smaller energy band gaps than silicon, such as InSb and InAs, it is expected that the energy band gap disappears at smaller magnetic fields. Namely, the critical magnetic field is expected to be smaller than $2.22 \times 10^4$ (T) for materials with smaller energy band gaps. Calculations for such materials will be the subject of future work.

The effect of the electron-electron interaction is not considered in the present calculations. For more accurate description of the second-order phase transition, the effect of electron-electron interaction should be taken into account. For example, the current-density functional theory can be used for this aim [22-24]. It seems to be interesting to investigate how the electron-electron interaction changes the energy band structure of silicon immersed in a magnetic field and to what extend it makes revisions to the second-order phase transition phenomena in the present case.

**Acknowledgements**

This work was partially supported by Grant-in-Aid for Scientific Research (No. 18K03510 and No. 18K03461) of the Japan Society for the Promotion of Science.

**Data Availability**

The data that support the findings of this study are available upon reasonable request from the authors.

**Figure captions**

Figure 1: Magnetic-field dependences of the magnetization and energy band gap. The left vertical axis denotes the magnetization that is calculated by the derivative of the total energy with respect to the magnetic field. The right vertical axis denotes the energy band gap of the crystalline silicon immersed in a magnetic field.

Figure 2: The energy band structure $E(\mathbf{k})$ for $p/q = 1/151$ that corresponds to $B = 743.1$ (T). The $\Gamma$ and $R_x$ points correspond to $\bar{\mathbf{k}} = (0,0,0)$ and $\bar{\mathbf{k}} = (1,0,0)$, respectively. Energy bands are nearly flat in the $k_x - k_y$ plane that is perpendicular to the magnetic field.

Figure 3: Energy band structures $E(\mathbf{k})$ for (a) $p/q = 14/101$ that corresponds to $B = 9.9991 \times 10^3$ (T), (b) $p/q = 14/101$ that corresponds to $B = 1.5554 \times 10^4$ (T), (c) $p/q = 31/157$ that corresponds to $B = 2.2156 \times 10^4$ (T) and (d) $p/q = 20/101$ that corresponds to $B = 2.2220 \times 10^4$ (T), respectively.

Figure 4: Energy band structures $E(\mathbf{k})$ for (a) $p/q = 13/59$ that corresponds to $B = 2.4725 \times 10^4$ (T), (b) $p/q = 25/113$ that corresponds to $B = 2.4826 \times 10^4$ (T), (c) $p/q = 35/157$ that corresponds to $B = 2.5015 \times 10^4$ (T) and (d) $p/q = 24/107$ that corresponds to $B = 2.5169 \times 10^4$ (T), respectively.

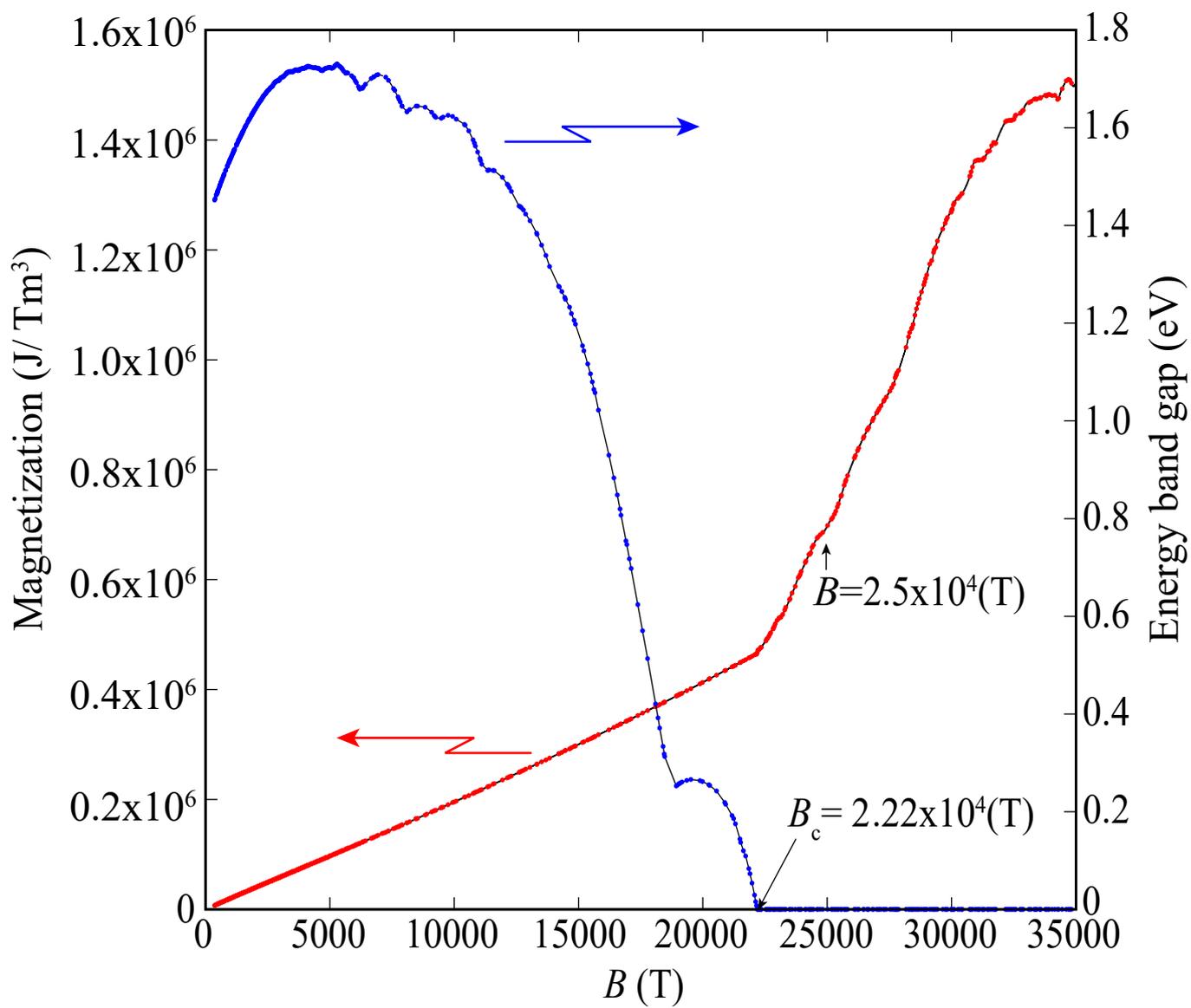

Figure 1

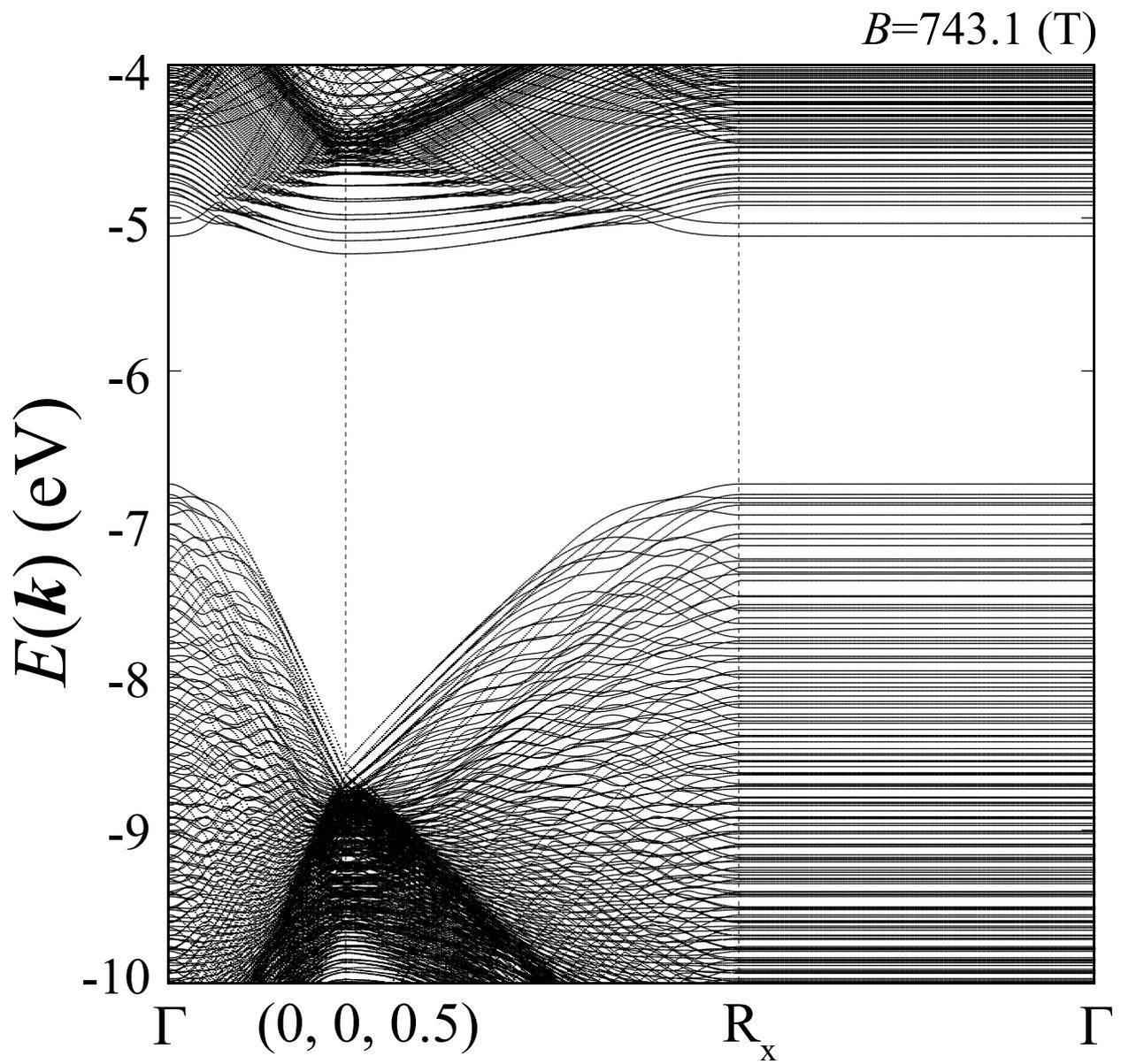

Figure 2

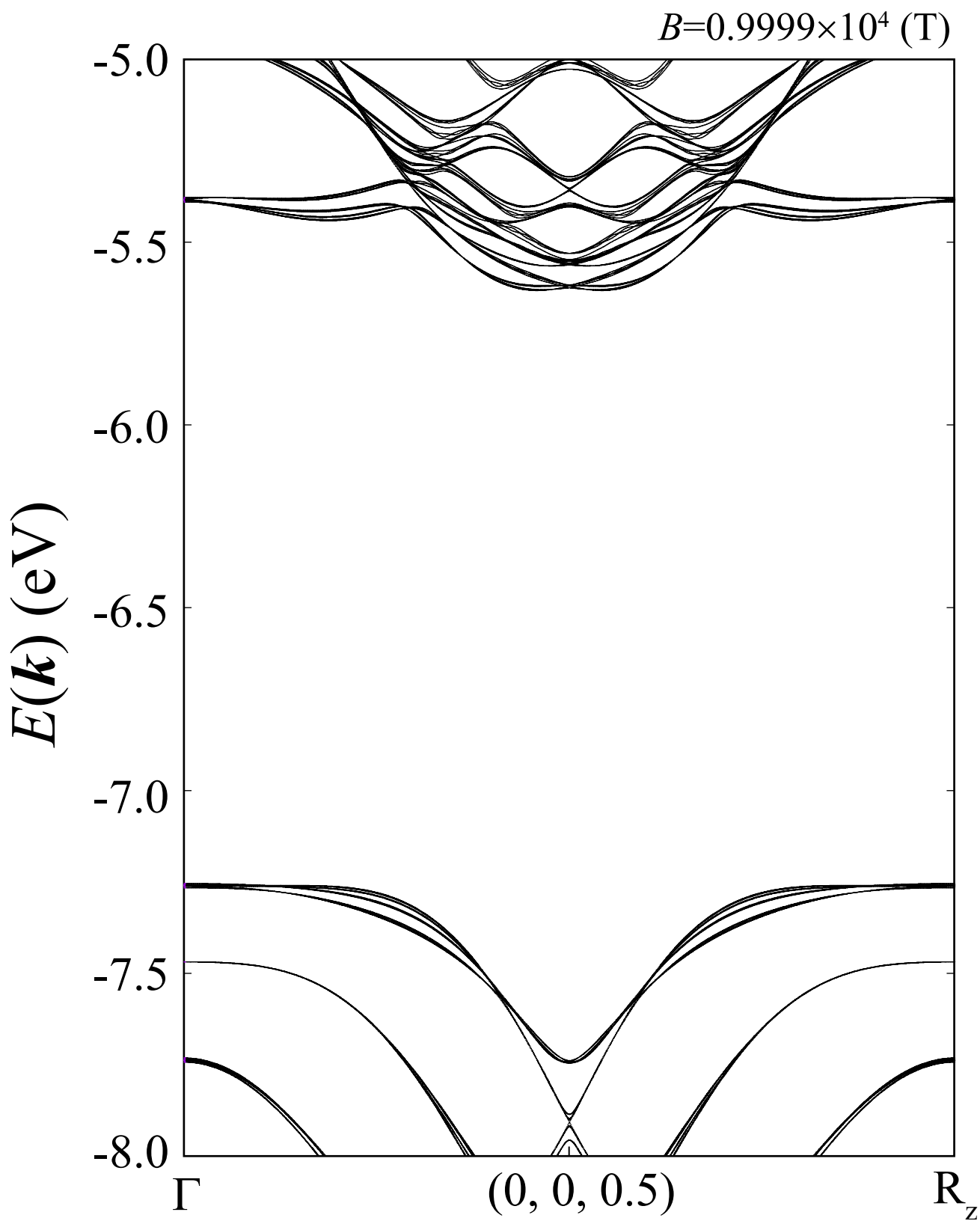

Fig. 3(a)

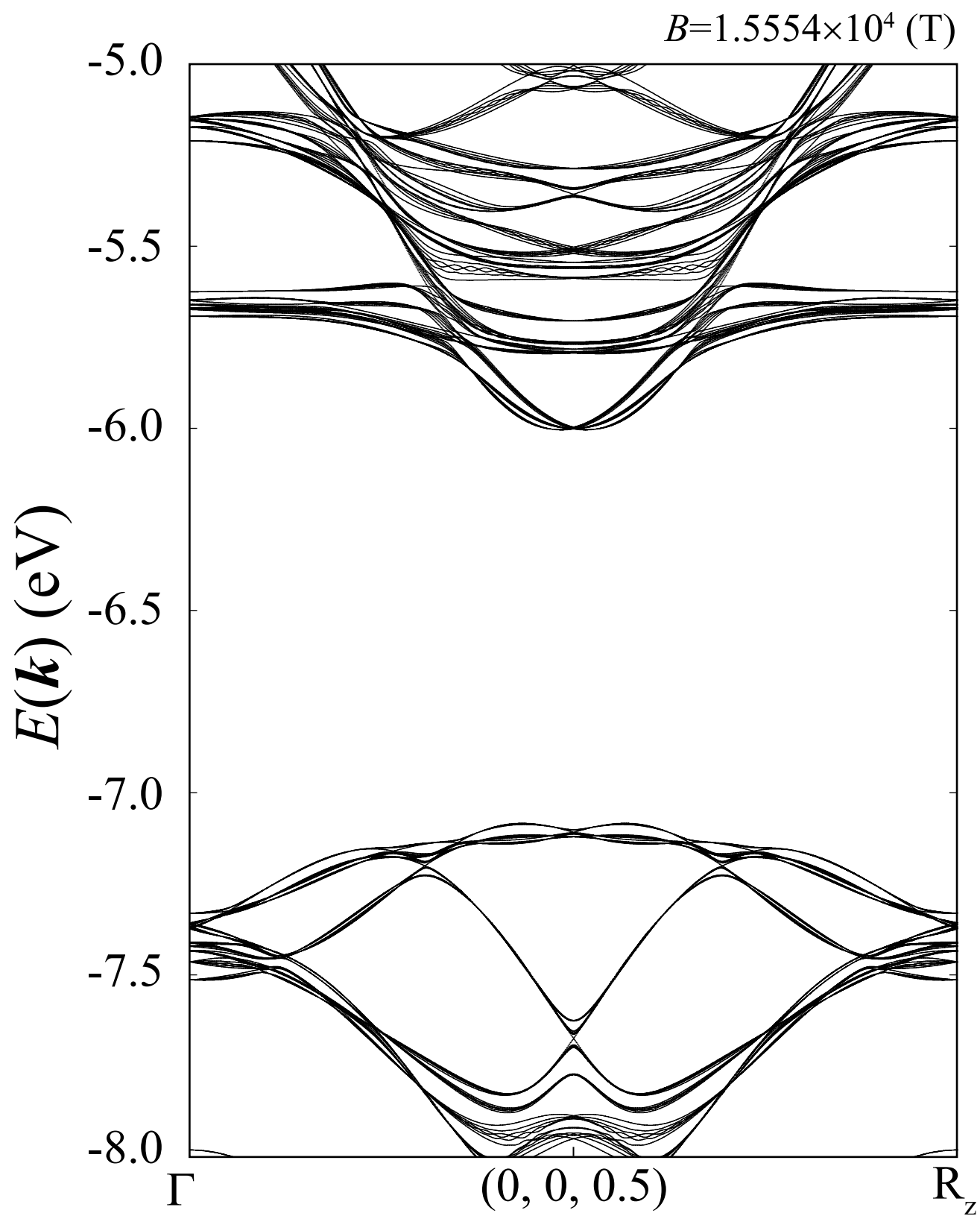

Fig. 3(b)

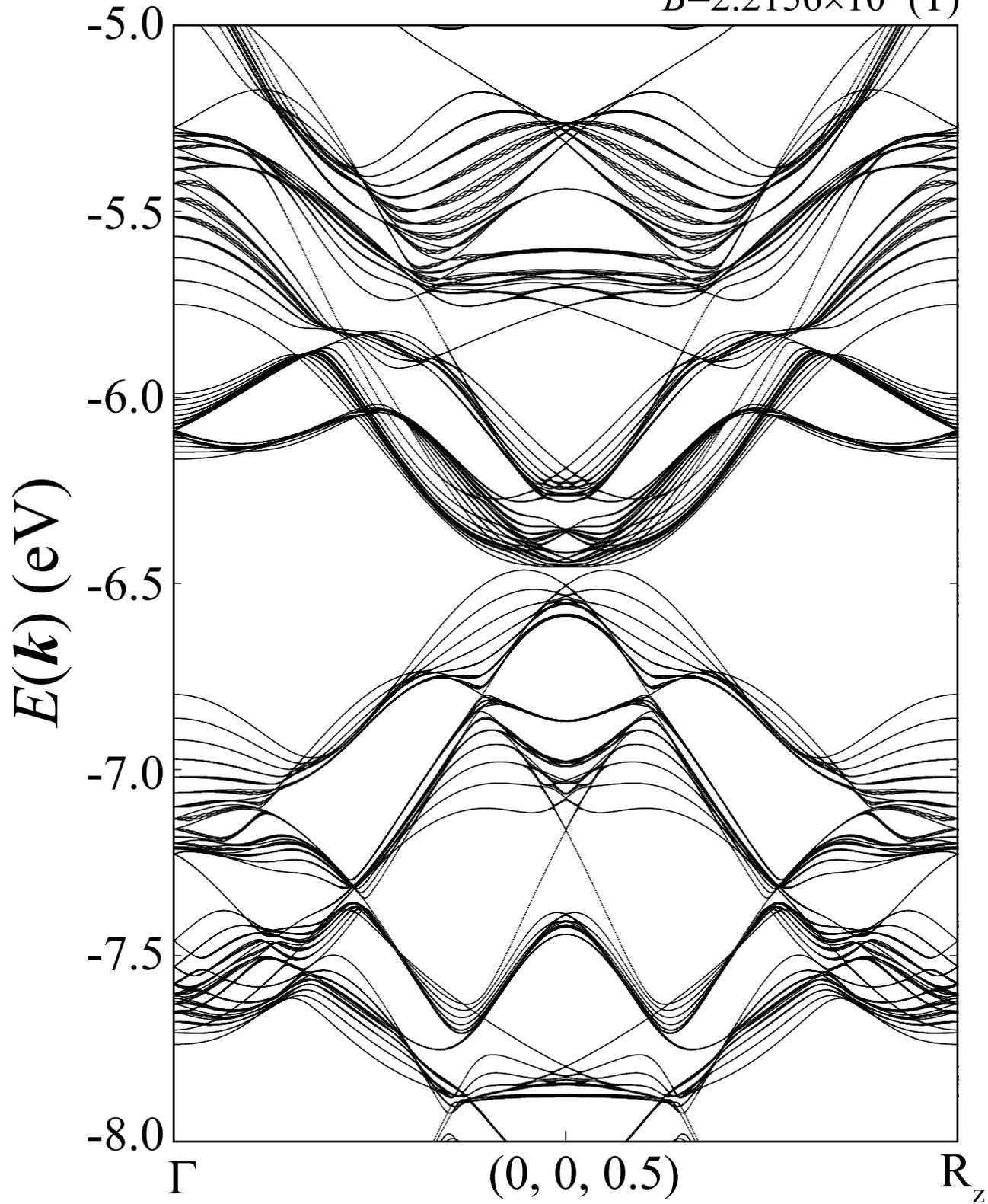

Fig. 3(c)

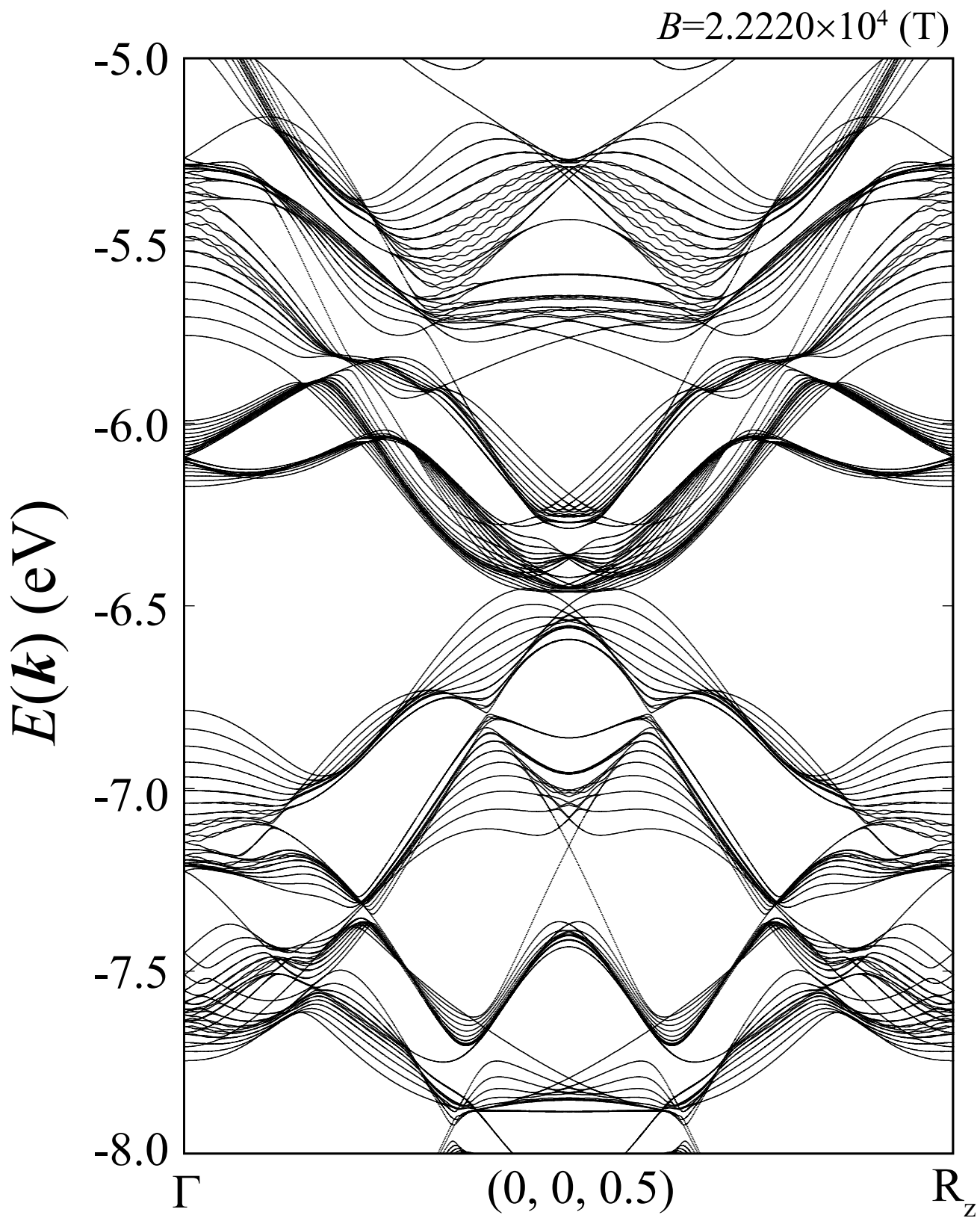

Fig. 3(d)

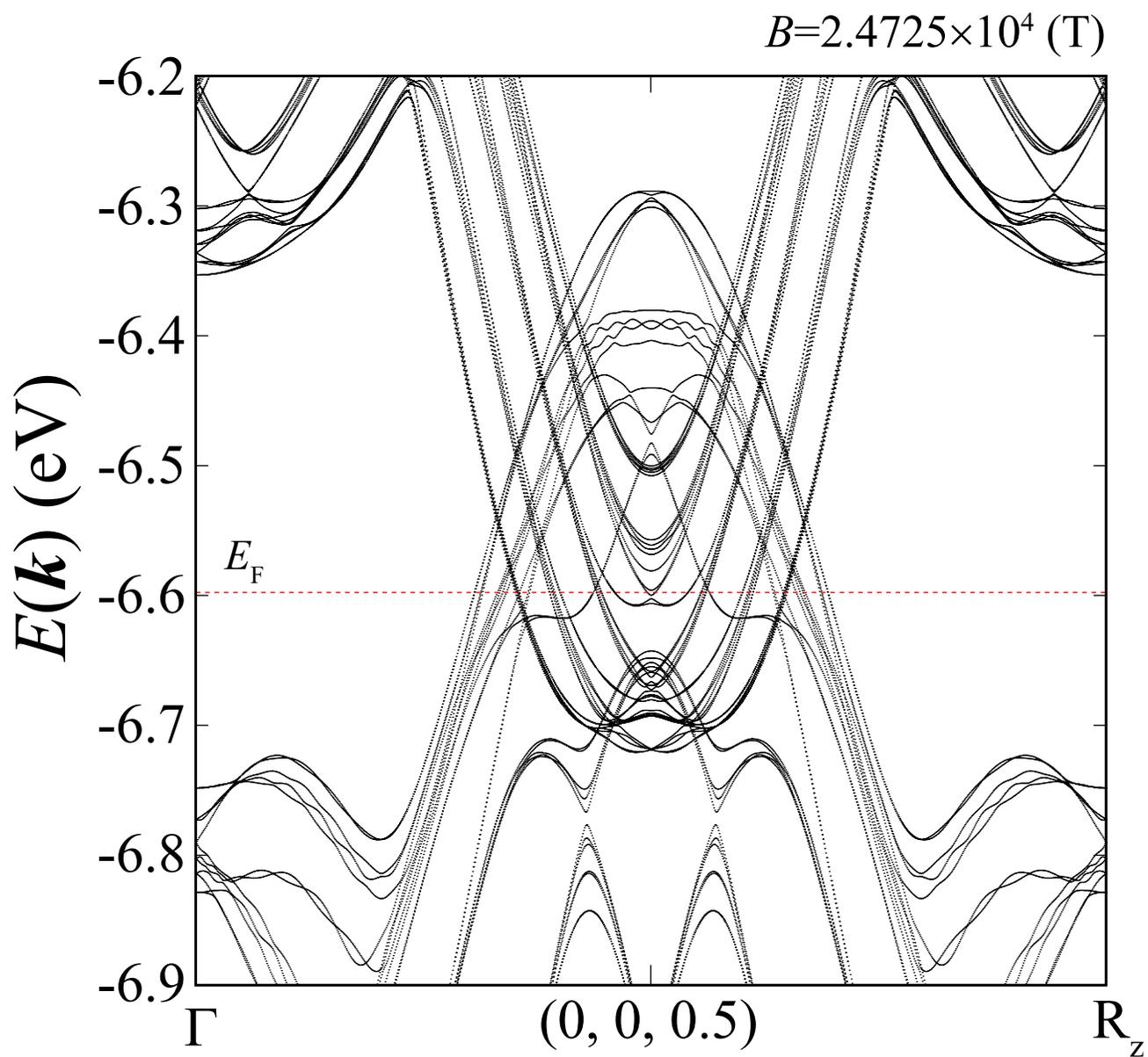

Fig. 4(a)

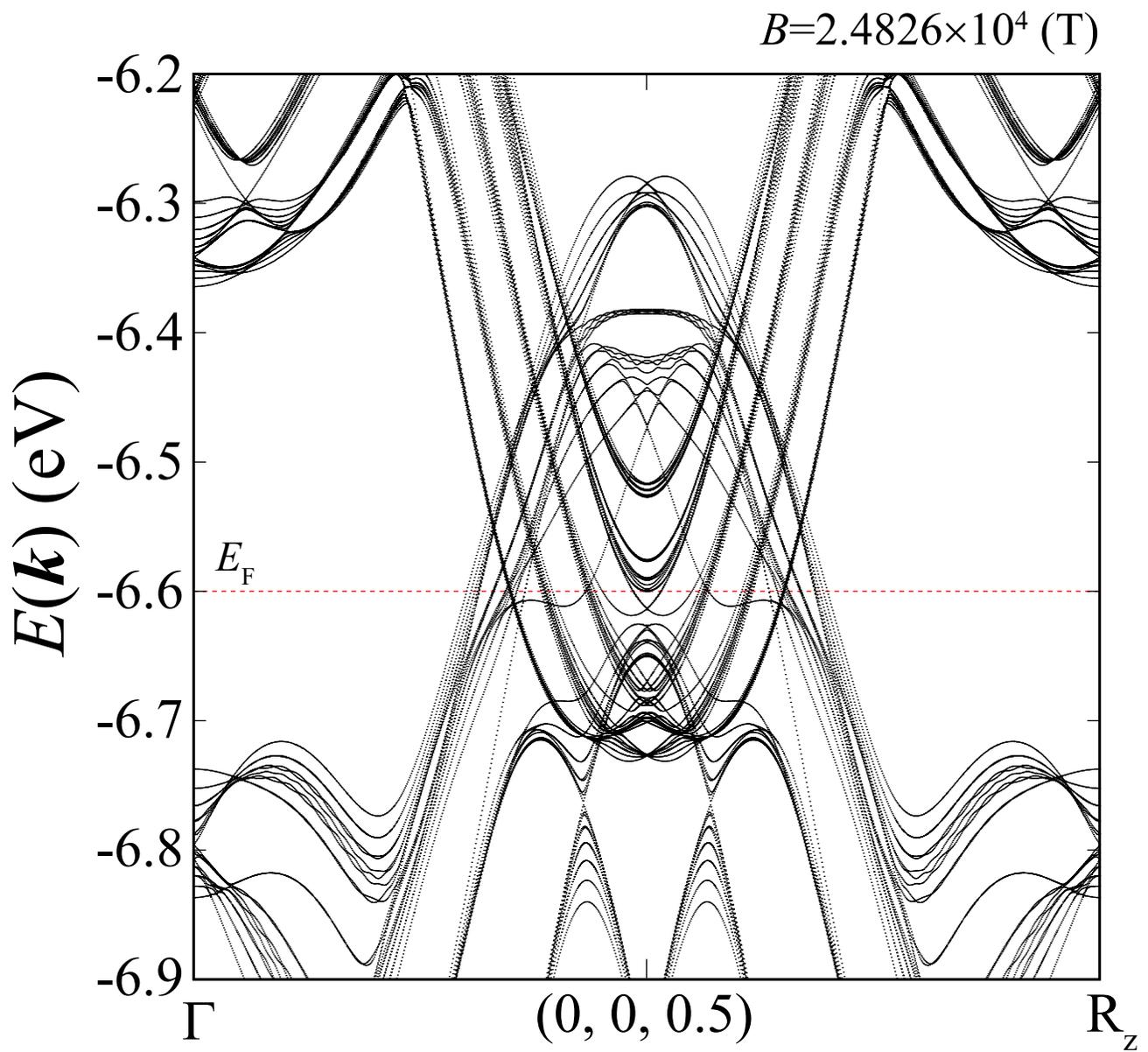

Fig. 4(b)

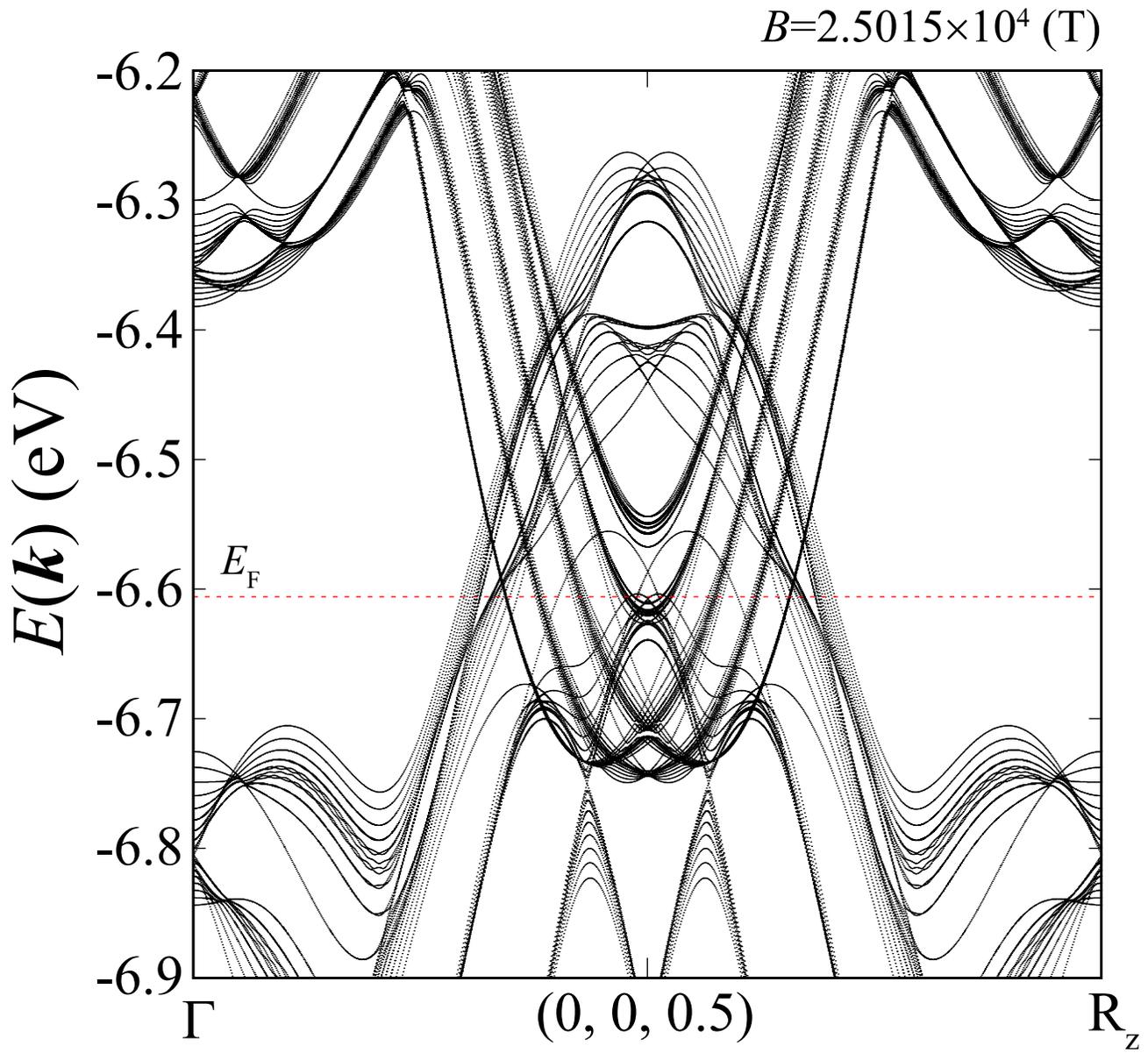

Fig. 4(c)

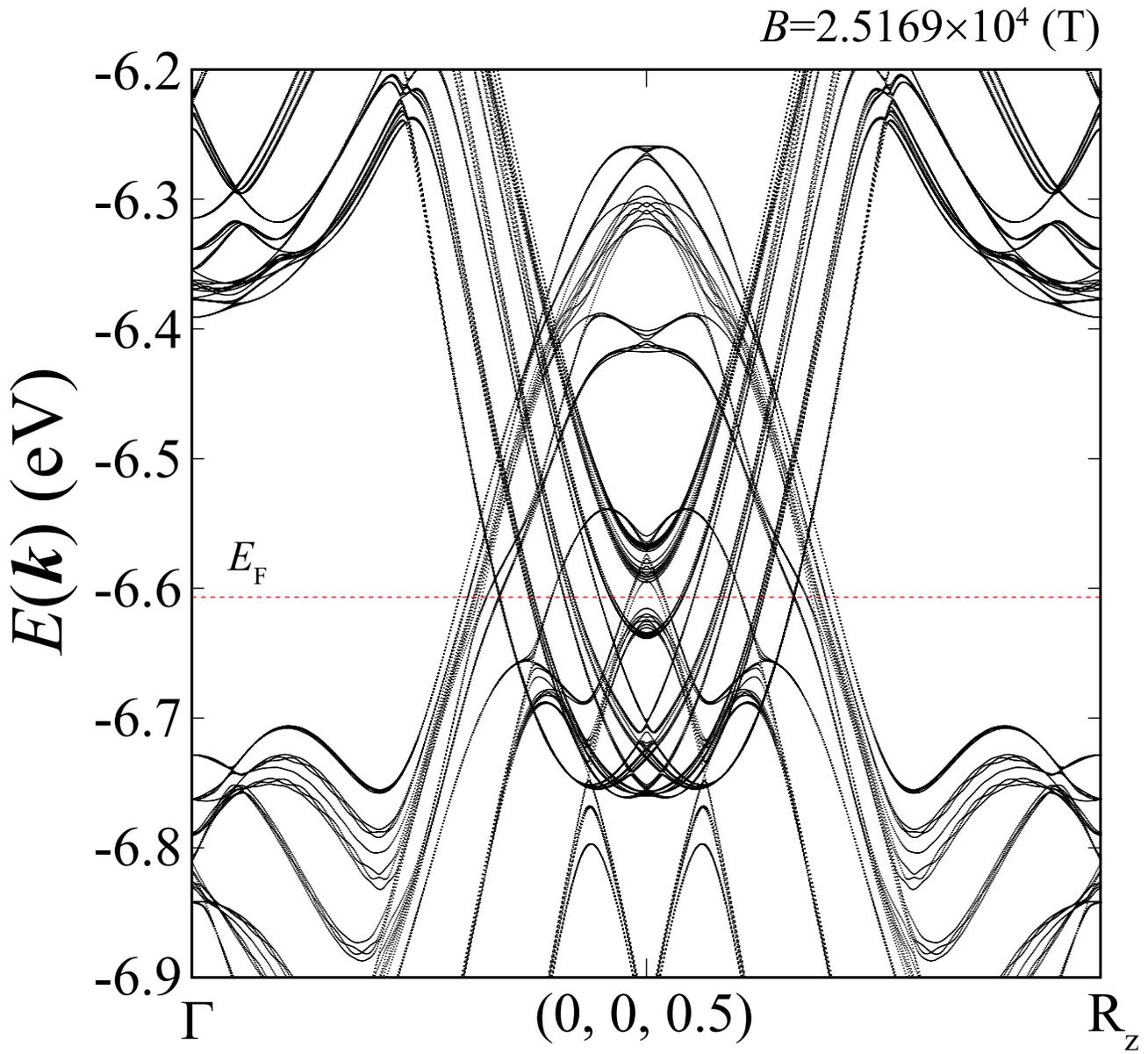

Fig. 4(d)